\documentclass[a4paper,aps,11pt,nofootinbib]{revtex4}
\usepackage{amsmath}
\usepackage{amssymb}
\usepackage{amsfonts}
\usepackage{color}
\usepackage{comment}
\usepackage{ascmac}
\usepackage{setspace}
\usepackage{color}
\definecolor{DarkBlue}{rgb}{0,0,0.7}

\definecolor{DarkRed}{rgb}{0.65,0,0}

\definecolor{DarkGreen}{rgb}{0,0.3,0}

\begin{document}

%%%%%%%%%
\def\N{{\cal N}^3}
%%%%%%%%%

\title{G\"odel-type Solutions in Einstein-Maxwell-Scalar Field Theories}

%%%% To generate auto affiliation numbers please use \author{}\affil{} command

\begin{spacing}{1.0}
\hfill{OCU-PHYS 548}

\hfill{AP-GR 173}

\hfill{NITEP 120}
\end{spacing}

\author{Hideki Ishihara}
\email{ishihara@osaka-cu.ac.jp}
\author{Satsuki Matsuno}
\email{sa21s010@osaka-cu.ac.jp}
\affiliation{
 Department of Mathematics and Physics,
 Graduate School of Science, 
 Nambu Yoichiro Institute of Theoretical and Experimental Physics (NITEP),
 Osaka City University,
 Osaka 558-8585, Japan
}

%%% To include the collaborator name... Please use the command "\collaborator"
%%% For example: \collaborator{ATLAS Collaboration}

\begin{abstract}%
We show that one-parameter family of G\"odel-type metrics are exact solutions in a 
wide class of Einstein-Maxwell-scalar field theories. 
In these solutions, the gauge field has a 1-form expectation value parallel to 
the timelike contact form that represents the rotation of the G\"odel universe.
\end{abstract}

\maketitle

\section{Introduction}

The cosmological metric found by G\"odel, so-called G\"odel universe \cite{Godel(1949)}, is 
one of the most intriguing exact solutions to the Einstein equations. 
Striking properties of the metic are the homogeneous geometry in space and time, 
the cosmological rotation, and the appearence of closed timelike curves 
(CTCs) through each points.  
For a long period after the pioneering work by G\"odel, much attention is paid 
on many G\"odel-type solutions with a cosmological constant and 
fluids in general relativity \cite{Som-Raychaudhuri, 
Banerjee-Banerji, Bampi-Zordan, Reboucas, Raychaudhuri-Thakurta, 
Reboucas-Tiomno, Reboucas-Teixeira, Reboucas-Aman-Teixeira, Dunn}
 (see also references therein). 
%, Reboucas-Tiomno, Goedel_type2}. 

The G\"odel-type spacetime is a direct product of a 3-dimensional space 
with a Lorentzian metric and a 1-dimensional space. 
The 3-dimensional spacetimes make a one-parameter family of squashed 
3-dimensional anti-de Sitter spaces (AdS$^3$) \cite{Rooman:1998xf}. 
Such a 3-dimensional spacetime is a contact space with a timelike contact form 
that describes the rotation of the spacetime \cite{Duggal}. 
This structure is easily extended to higher-dimensional theories,  
then, recently, the G\"odel-type metrics in higher dimensions are studied in the context of 
supersymmetric theories \cite{Barrow-Dabrowski, Gauntlett_etal, 
Harmark-Takayanagi}.  

We show, in this paper, that the one-parameter family of G\"odel-type metrics 
are exact solutions to the Einstein equations couples to purely field theories 
that consist of a U(1)-gauge field, a real scalar field, and a complex scalar field 
with a wide class of potentials. 
The interesting property of the solutions 
is that the gauge field has a 1-form expectation value parallel to the 
timelike contact form.   
Owing to the non-vanishing 1-form expectation value, 
the complex scalar field has an expectation value that is shifted off the 
potential minimum in these solutions. 
These solutions provide new non-trivial symmetry-breaking states 
in simple models, then  
it suggests that a variety of gauge field theories admit the solutions 
with the G\"odel-type metrics. 
As is seen later, for example, a field theory admits two G\"odel-type solutions: 
one is with CTCs, and the other is without CTC.  
It would open the possibility of phase transition from the former state to 
the latter state. 

\bigskip

%%%%%%%%%%%%%%%%%%%%%%%%%%%
\section{Basic Model}
%%%%%%%%%%%%%%%%%%%%%%%%%%%
We consider the action
\begin{align}
	S=\int \sqrt{-g}d^4x &\left(\frac{1}{16\pi G} R
		-g^{\mu\nu}(\nabla_{\mu}\Phi)^{\ast}(\nabla_{\nu}\Phi)
		-V(\Phi^*\Phi)
		-\frac12 g^{\mu\nu}(\partial_{\mu}\Psi)(\partial_{\nu}\Psi)
		-\frac{1}{4}F_{\mu\nu}F^{\mu\nu}
	\right),
\label{eq:action}
\end{align}
where $R$ is the scalar curvature with respect to a metric $g_{\mu\nu}$, 
$\Phi$ is a complex scalar field, $\Psi$ is a real scalar field,
and $F_{\mu\nu}$ is the field strength of a U(1) gauge field $A_\nu$. 
The complex scalar field has a potential $V(\Phi^*\Phi)$, and
couples with the gauge field through $\nabla_{\nu}\Phi :=(\partial_{\nu}-ieA_\nu)\Phi$.

By variation of the action \eqref{eq:action}, we derive field equations:
\begin{align}
	&\frac{1}{\sqrt{-g}}\nabla_\mu(\sqrt{-g} g^{\mu\nu} \nabla_\nu \Phi)
	-\frac{\partial V}{\partial\Phi^*}=0,
 \label{eq_phi}
 \\
	&\frac{1}{\sqrt{-g}}\partial_\mu(\sqrt{-g} g^{\mu\nu}\partial_\nu \Psi)=0,
 \label{eq_psi}
 \\
	&\frac{1}{\sqrt{-g}}\partial_\mu(\sqrt{-g} F^{\mu\nu})= J^{\nu},
\label{Maxwell_eq}
\\
	&R_{\mu\nu}=8\pi G\left( T_{\mu\nu}-\frac12 g_{\mu\nu} T_\alpha^\alpha\right) ,
\label{Einstein_eq}
\end{align} 
where $J_{\mu}$ is the electric current defined by
\begin{align}
  	J_{\mu} &:= ie\Big(\Phi^{\ast}\nabla_{\mu}\Phi-\Phi(\nabla_{\mu}\Phi)^{\ast}\Big), 
\label{eq:current}
\end{align} 
and the energy-momentum tensor $T_{\mu\nu}$ is given by
\begin{align}
T_{\mu\nu}
	=& 2(\nabla_{\mu}\Phi)^{\ast}(\nabla_{\nu}\Phi)
	-g_{\mu\nu}\left(g^{\alpha\beta}(\nabla_{\alpha}\Phi)^{\ast}(\nabla_{\beta}\Phi)+V(\Phi^*\Phi)\right)
\cr
	&+(\partial_{\mu}\Psi)(\partial_{\nu}\Psi)
	-\frac12 g_{\mu\nu} g^{\alpha\beta}(\partial_{\alpha}\Psi)(\partial_{\beta}\Psi)
\cr
	&+ F_{\mu\alpha}F_{\nu}^{~\alpha}
	-\frac{1}{4}g_{\mu\nu}F_{\alpha\beta}F^{\alpha\beta}.
\label{eq:T_munu}
\end{align}

%\subsection{Metric}

We assume the total spacetiem ${\cal M}^4$ is a direct product of a 3-dimensional 
spacetime $\N$ and $\mathbb R^1$, where the metric takes the form
\begin{align}
	ds_{\cal M}^2 = ds_{\cal N}^2 + d\chi^2. 
\label{metric}
\end{align}
The metric on $\N$ is assumed to be 
\begin{align}
	ds_{\cal N}^2 = -(dt + f(r) d\theta)^2
 			+ dr^2 + h^2(r) d\theta^2, 
\label{metric_N}
\end{align}
where the coordinate $\theta$ is an angular coordinate with the period $2\pi$. 
The functions $f(r)$ and $h(r)$ 
should be determined later.

We introduce the 1-form basis, 
\begin{align}
	\sigma^0= dt+ f(r) d\theta, \quad \sigma^1= d r ,\quad 
	\sigma^2= h d\theta, \quad \sigma^3= d\chi, 
\label{basis}
\end{align}
so as to rewrite the metric \eqref{metric} as
\begin{align}
	ds^2=-(\sigma^0)^2 + (\sigma^1)^2 + (\sigma^2)^2 + (\sigma^3)^2.  
\label{metric_form}
\end{align}
We note that  
\begin{align}
	d\sigma^0 
	= \frac{f'}{h} \sigma^1\wedge \sigma^2, 
\end{align}
where the prime denotes the derivative with respect to $r$. Therefore, if $f'\neq 0$, 
\begin{align}
	\sigma^0\wedge d\sigma^0 \neq 0 
\label{contact_form}
\end{align}
holds. 
The 1-form $\sigma^0$ satisfying \eqref{contact_form} is called a contact form 
on the 3-dimensional Lorentzian space $\N$. 

%\subsubsection{Ricci curvature}

Furthermore, if we assume 
\begin{align}
	f'= 2\omega h, \quad (\omega : const.),
\end{align}
it is known that $\N$ is a quasi-Sasaki space with a Lorentzian metric\cite{Takahashi,Blair}. 
The Ricci curvature tensor with respect to the metric \eqref{metric_form} reduces 
to the simple form: 
\begin{align}
	R_{ab}=2\omega^2\sigma^0_a \sigma^0_b 
	+ \left(2\omega^2-\frac{ h''}{h}\right)(\sigma^1_a\sigma^1_b +\sigma^2_a\sigma^2_b).  
\label{Ricci}
\end{align}

\bigskip

%%%%%%%%%%%%%%%%%%%%%%%%%%%%%%%%%%%%%%%%%%%%%%%%%%%%%%%%%%%%%
\section{ Equations of scalar fields and Maxwell field}
%%%%%%%%%%%%%%%%%%%%%%%%%%%%%%%%%%%%%%%%%%%%%%%%%%%%%%%%%%%%%

We assume the scalar fields and the gauge field as 
\begin{align}
	\Phi=v, \quad
	\Psi = \Psi(\chi),  \quad
	A=B \sigma^0,
\label{assumption}
\end{align}
where $v$ and $B$ are assumed to be non-vanishing real positive constants. 
The equation \eqref{eq_phi} of $\Phi$ reduces to
\begin{align}
	\frac{1}{2}\frac{d V(v)}{dv}= e^2 B^2 v ,
\label{eq_Phi}
\end{align}
where $V(v)$ is defined by replacement $\Phi^*\Phi$ by $v^2$ in $V(\Phi^*\Phi)$. 
Owing to the non-vanishing gauge field, the expectation value 
$v$ of $\Phi$ is shifted off the stationary points of the potential $V(v)$. 

The equation \eqref{eq_psi} of $\Psi$ reduces to
\begin{align}
	\partial_\chi^2 \Psi(\chi)=0,
\end{align} 
then by using a real constant $k$, we have
\begin{align}
	\Psi = k \chi+const. 
\label{sol_Psi}
\end{align}

%\subsubsection{Maxwell eq}

The Maxwell equation \eqref{Maxwell_eq} in the differential form is written as
\begin{align}
	\star(d{\star F}) =-J, 
\label{Maxwell_diff_form}
\end{align}
where $\star$ denotes the Hodge dual operator. 
The current \eqref{eq:current} carried by $\Phi$ reduces to
\begin{align}
  	J &= 2 e^2 v^2 A . 
\label{J}
\end{align}
Then, \eqref{Maxwell_diff_form} becomes
\begin{align}
	\star (d{\star F})=- 2 e^2 v^2 A.
\label{Proca_eq}
\end{align}
Owing to the scalar field $\Phi$ has the expectation value $v$, 
the gauge field acquires the mass $2e^2 v^2$, then 
we have the Proca equation \eqref{Proca_eq}. 
On the other hand, from the assumptions \eqref{assumption} that the gauge 1-form 
$A$ is parallel to the contact form $\sigma^0$, 
the field strength $F$ is 
\begin{align}
	&F=dA =2\omega B \sigma^1\wedge \sigma^2,
\label{F}
\end{align}
then the left hand side of \eqref{Maxwell_diff_form} becomes
\begin{align}
	\star (d{\star F})=-4\omega^2 A.
\label{Proca_eq_2}
\end{align}
Therefore, \eqref{Proca_eq} requires
\begin{align}
	\omega^2=\frac12 e^2v^2. 
\label{omega}
\end{align}

\bigskip

%%%%%%%%%%%%%%%%%%%%%%%%%%%%%%%%%%%%%%%%%%%%%%%%%%%%%%%%%%%%%
\section{Einstein's equations}
%%%%%%%%%%%%%%%%%%%%%%%%%%%%%%%%%%%%%%%%%%%%%%%%%%%%%%%%%%%%%

The total energy-momentum tensor \eqref{eq:T_munu} in the 1-form basis \eqref{basis} 
is given by $T_{ab}=T_{ab}^\Phi+T_{ab}^\Psi+T_{ab}^F$, where 
\begin{align}
	&T_{ab}^\Phi
		= \left(e^2 B^2v^2+ V(v)\right)\sigma^0_a\sigma^0_b 
		+ \left(e^2 B^2 v^2	- V(v) \right)
		 (\sigma^1_a\sigma^1_b+ \sigma^2_a\sigma^2_b+ \sigma^3_a\sigma^3_b ),
\label{T_Phi}\\
	&T_{ab}^{\Psi} 
	=\frac{k^2}{2}(\sigma^0_a \sigma^0_b -\sigma^1_a \sigma^1_b 
	-\sigma^2_a \sigma^2_b+\sigma^3_a \sigma^3_b),
\label{T_Psi}
\\
	&T_{ab}^{F}
	=2\omega^2 B^2(\sigma^0_a \sigma^0_b +\sigma^1_a \sigma^1_b 
		+\sigma^2_a \sigma^2_b-\sigma^3_a \sigma^3_b).
\label{T_F}
\end{align}

%\subsection{Ricci curvature}

From \eqref{Ricci} and \eqref{T_Phi}-\eqref{T_F}, 
the Einstein equation \eqref{Einstein_eq} reduces to
\begin{align}
	2\omega^2\sigma^0_a\sigma^0_b 
	&+ \left(2\omega^2 -\frac{h''}{ h}\right)(\sigma^1_a\sigma^1_b +\sigma^2_a\sigma^2_b)  
\cr
	=& \left(2e^2 B^2v^2 -V(v)+ 2\omega^2 B^2\right)\sigma^0_a\sigma^0_b 
	+ \left(V(v) +2\omega^2 B^2\right)
	 (\sigma^1_a\sigma^1_b+ \sigma^2_a\sigma^2_b)
\cr
	&\hspace{1cm}
	+ \left(V(v) + k^2 -2\omega^2 B^2\right) \sigma^3_a \sigma^3_b, 
\label{Einstein_eq_2}
\end{align}
where we use the unit $8\pi G=1$. Namely, we have a set of coupled equations: 
\begin{align}
	&	2\omega^2=2e^2 B^2v^2 -V(v)+ 2\omega^2 B^2,
\label{R00}
\\
	&	2\omega^2 -\frac{h''}{h} =V(v) +2\omega^2 B^2,
\label{R11}
\\
	&0=	V(v)+ k^2-2\omega^2 B^2 . 
\label{R33}
\end{align}

%\subsection{solution}

From \eqref{R00}, \eqref{R33} with \eqref{omega}, for non-vanishing $v$, we have
\begin{align}
	&B^2 =\frac12-\frac{k^2}{2e^2v^2} 
		=\frac1{3e^2v^2} (e^2v^2+V(v)),
\label{B2}\\
	&k^2=\frac13 (e^2v^2 -2V(v)) , 
\label{k2}
\end{align}
and \eqref{eq_Phi} becomes
\begin{align}
	3v\frac{dV(v)}{dv}-2e^2v^2 -2V(v) =0. 
\label{eq_v}
\end{align}
For a given potential $V(\Phi^*\Phi)$, let $v_0$ be a root of the algebraic 
equation \eqref{eq_v}. 
Inserting $v=v_0$ into \eqref{B2} and \eqref{k2}, we express $B^2$ and $k^2$ 
by $v_0$. 

Using  \eqref{omega}, \eqref{B2} and \eqref{k2} we rewrite \eqref{R11} as
\begin{align}
 	h''=2 k^2 h. 
\end{align}
For regularity, $h$ should behave $h\to r$ as $r \to 0$, then we have 
\begin{align}
	h= \frac{1}{\sqrt{2} k } \sinh\!{\sqrt{2} k r},
\quad
	f= \frac{2\omega}{k^2} \sinh^2\!{\frac{k r}{\sqrt{2}}}. 
\end{align}
Therefore, we obtain G\"odel-type metrics: 
\begin{align}
	ds^2 &=-\left(dt + \frac{2\omega}{k^2}\sinh^2\!{\frac{k r}{\sqrt{2}}} ~d\theta\right)^2 
 	+dr^2 + \frac{1}{2k^2}\sinh^2\!{\sqrt{2}k r}~ d\theta^2 
 	+ d\chi^2 ,
\label{Godel_metric}
\end{align}
where the parameters $\omega$ and $k$ are given by \eqref{omega} and \eqref{k2} for $v=v_0$. 
The metric \eqref{Godel_metric} is already discussed 
by Rebou\c{c}as and Tiomno \cite{Reboucas-Tiomno}. 
%\subsubsection{Closed timelike curves}

From the explicit form of the metric \eqref{Godel_metric}, 
we see that if $2\omega^2/k^2 > 1$ closed timelike curves (CTCs) appear 
in the region $r_c < r$, where the critical radius $r_c$ is given by
\begin{align}
	\sinh^2{ \frac{k r_c}{\sqrt{2}}} 
		= \left(\frac{2\omega^2}{k^2}-1\right)^{-1}. 
\end{align} 
A circle of a radius $r=const. >r_c$ on a $t=const.$ surface is a CTC. 
The radius of CTC becomes larger and diverges as $k^2$ approaches to 
$2\omega^2$. 

By use of \eqref{omega} and \eqref{k2} the existence condition of CTC becomes
\begin{align}
	V(v_0)>-e^2v_0^2. 
\label{CTC}
\end{align}
Positivity of $B^2$ and $k^2$ in \eqref{B2} and \eqref{k2} requires 
\begin{align}
	\frac12 e^2v_0^2 > V(v_0) > -e^2v_0^2 , 
\end{align}
where the upper bound corresponds to $k \to 0$ and the lower to $B\to 0$. 
Then, if $B\neq 0$, CTCs appear generally. 
In the special case $B = 0$, from \eqref{R00} and \eqref{R33} we have $k^2=2\omega^2$, 
and \eqref{Godel_metric} reduces to a metric of AdS$^3\times \mathbb R^1$ in the form
\begin{align}
	ds^2 &=\frac{1}{2k^2}\left(-\left(d\tau + 2\sinh^2\!{\frac{\rho}{2}} ~d\theta\right)^2 
 		+d\rho^2 + \sinh^2\!\rho ~d\theta^2 \right) + d\chi^2 ,
\label{AdS_metric}
\end{align}
where $\rho:=\sqrt{2} kr$ and $\tau := \sqrt{2} kt$. In this case, there is no CTC. 

\bigskip

%%%%%%%%%%%%%%%%%%%%%%%%%%%%%%%%%%%%%%%%%%%%%%%%%%%%%%%%%%%%%
\section{ Examples}
%%%%%%%%%%%%%%%%%%%%%%%%%%%%%%%%%%%%%%%%%%%%%%%%%%%%%%%%%%%%%

Here, we present two examples of the potential $V(\Phi^*\Phi)$.  
The first case is a massive free complex scalar field $\Phi$ with a negative vacuum energy. 
The potential is given by 
$V(\Phi^*\Phi)=m^2\Phi^*\Phi - V_0$, where $m$ and $V_0$ are positive constants.  
Solving \eqref{eq_v}, we have 
\begin{align}
	v_0^2=\frac{V_0}{e^2-2m^2}, 
\end{align}
where $e^2>2m^2$ should be assumed. From \eqref{B2} and \eqref{k2}, we have
\begin{align}
	B^2=\frac{m^2}{e^2}, \quad k^2=\frac13 V_0, 
\end{align}
and the metric given by \eqref{Godel_metric}. 

In the special case $e^2=2m^2$ and $V_0=0$, 
\eqref{eq_v} allows an arbitrary value of $v_0$, and we have
\begin{align}
	B^2 =\frac12, \quad
	k^2 =0, 
\end{align}
and the metric becomes the one found by Som and Raychaudrhi \cite{Som-Raychaudhuri} 
in the form
\begin{align}
	ds^2&=-\left(dt + \omega r^2  d\theta\right)^2 
 		+dr^2 + r^2  d\theta^2 + d\chi^2 ,
\end{align}
where $\omega^2=e^2 v_0^2/2$ can take an arbitrary value.

Next, we consider the potential of the form  
$V(\Phi)= \frac{\lambda}{2}\left(\Phi^*\Phi-\eta^2\right)^2-e^2\eta^2$, where $\eta$ is 
a constant. 
Solving \eqref{eq_v}, we have two cases:
\begin{align}
	{\rm(i)}\quad v_0^2=\eta^2, \qquad 
	{\rm(ii)}\quad v_0^2=\frac{1}{5\lambda}(2e^2-\lambda \eta^2). 
\end{align}
In the case (i), we obtain
\begin{align}
	B^2=0, 
\quad
	k^2=e^2\eta^2 , 
\end{align}
and the metric of AdS$^3\times \mathbb R^1$ given by \eqref{AdS_metric}.

In the case (ii), we have
\begin{align}
	B^2=\frac{2}{5e^2}(e^2- 3\lambda\eta^2), 
\quad
	k^2	=\frac{1}{25\lambda} \left(2e^4 +23 e^2 \lambda\eta^2  - 12 \lambda^2\eta^4\right). 
\end{align}
If $e^2 = 3\lambda\eta^2$, the case reduces to the case (i), 
and if $e^2 > 3\lambda\eta^2$, 
%we have $v_0^2 > \eta^2$, and 
the metric takes the form of \eqref{Godel_metric}. 

In this example, we see that there are two non-trivial solutions  
in a theory with the potential $V$. 
One is a G\"odel-type with a non-vanishing 1-form expectation value of 
the gauge field, and the other is AdS$^3 \times \mathbb R^1$ with the vanishing gauge field. 
The former has CTCs, while the latter does not. 

\bigskip

%%%%%%%%%%%%%%%%%%
\section{Summary}
%%%%%%%%%%%%%%%%%%
We have shown that 
the one-parameter family of G\"odel-type metrics 
are exact solutions to the coupled sysytem that consist of a U(1)-gauge field, 
a real scalar field, a complex scalar field with a wide class of potentials, 
and Einstein gravity. 
The solutions that have an expectation value of the complex scalar field and 
a 1-form expectation value of the gauge field. %The solutions have CTCs. 
We have presented a field theory that allowed two G\"odel-type solutions: 
metric with CTCs and metric without CTC. 
It seems interesting to study the possibility of transition 
between two symmetry-breaking states described by these solutions. 

\bigskip

\section*{Acknowledgment~~~} 

The authors thank Prof. Ken-ichi Nakao and Prof. Hiroshi Itoyama 
for valuable discussion.


%\bibliographystyle{ptephy}
%\bibliography{sample}
%
% once the .bbl file has been generated then place the text in your article.

%without this code before the command "\begin{thebibliography}{}" , an error will be %flagged. When the bibliography is provided as separate .bib file, then this code %should be placed above the commands "\bibliographystyle{}" and "\bibliography{}" %inside the main TeX file. 

\end{document}